\documentclass[aps, 10pt]{revtex4}
\usepackage{graphicx,subfigure}
\usepackage{epsfig}
\begin{document}
\title{Laughlin wave function, Berry Phase and quantisation}
\author{K. V. S. Shiv Chaitanya }
\email[]{ chaitanya@hyderabad.bits-pilani.ac.in}
\affiliation{Department of Physics, BITS Pilani, Hyderabad Campus, Jawahar Nagar, Shamirpet Mandal,
Hyderabad, India 500 078.}

\begin{abstract}
In this paper, we show that the Laughlin wave function is a Hamiltonian and its associated Berry connection as the Schr\"odinger equation by transforming the Schr\"odinger  equation into the  Kirchhoff equation which describes the evolution of $n$ point vortices in Hydrodynamics. This helps us to view 
the Berry connection associated with Laughlin wave function or Schr\"odinger  equation is not Hermitian, therefore we propose a self adjoint model Hamiltonian for the fractional quantum Hall effect, from the study of Kirchhoff equation, using superymmetric quantum mechanics whose solutions are complex Hermite polynomials.  The Schr\"odinger equation  as Berry connection allows us to formulate the problem of quantisation in terms of topology, as the quantum numbers are topological invariant which arise due to singularities in the  Kirchhoff equation. Quantisation arises as it continuously connects the topologically inequivalent Hamiltonians in the Hilbert space. 
\end{abstract}

\maketitle
\section{Introduction}
In recent years, the phases of condense matter, are characterized in terms of topology.   Topological insulators \cite{pd}, quantum Hall effect \cite{qh}, fractional quantum Hall effect \cite{fqh} are well-known.  The topological invariants don't change  from one shape to another shape under continuous deformation of geometrical object. In case of insulators it is characterized in terms of band gap, since the gap is large.  If the insulators are connected say from one insulator another insulator, without changing the band gap, that is, the system always remaining in ground state, they are called topologically equivalent insulators. In other words, when some parameter of the Hamiltonian is slowly changed, adiabatically, the ground state of the system remains unchanged.  The insulators which cannot be connected by the slowly changing Hamiltonian are called topologically inequivalent insulators. Connecting the topologically equivalent insulators gives rise to phase transition with the vanishing of gap. In this gapless state, topological invariants are quantized giving rise to current.

One of the well known examples of topological insulator is the quantum Hall effect  \cite{qh}. The integer quantum Hall effect was first discovered followed by fractional quantum Hall effect. In case of fractional quantum Hall effect, Robert Laughlin proposed the following wave function as an ansatz for the ground state wave function \cite{rbl}
\begin{equation}\label{lf}
\psi =  \prod_{N \geq i > j \geq 1}\left( z_i-z_j \right)^n  \exp\left( - \frac{1}{4l_B^2} \sum_i\vert z_i\vert ^2 \right)
\end{equation}
where, $z_i=x_i+iy_j$,  $N$ is number of electrons, $l_B=\sqrt{\frac{\hbar}{eB}}$ is magnetic length, $\hbar$ is Planck constant, $e$ is electric charge, $B$ is magnetic field, $\omega_B$ is cyclotron frequency $\omega_B=\frac{eB}{m}$, $z_k$ and $z_j$ are the position electrons for the ground state of a two-dimensional electron gas with the lowest Landau level, where $n$ is written in terms $\nu=1/n$ and  $n$ is an odd positive integer are filling numbers. As the Laughlin wave function is trial wave function, it is not an exact ground state of any potential, in particular not an exact ground state of Coulomb repulsion problem. But it has been tested numerically, for the Coulomb and several repulsive potentials, the Laughlin wave function has more than $99\%$ overlap with the true ground state \cite{dt}. The gap vanishes at the edges and the fractional charges are calculated using Berry's connection. In literature several model Hamiltonians are proposed which could admit Laughlin wave function as solution, first model of its kind was proposed by Haldane \cite{hal}.

The Berry connection for $N$ quasi-holes are given by \cite{dt}
\begin{equation}
A(\eta_j)=-\frac{i\nu}{2}\sum_{1\leq j\leq n,j\neq k}\frac{1}{\eta_k-\eta_j}+i\nu\frac{\bar{\eta_j}}{4l_B^2},\label{b1}
\end{equation}
here $\eta_i$ are the positions of the quasi holes and the $\nu$ are the filling numbers defined in equation (\ref{lf}) and  similar equation for $A(\bar{z_j})$ exist which describes the adiabatic transport of quasihole  at $z_i$ whan all other quasihole positions are fixed.

In this paper, we show that the Laughlin wave function (\ref{lf}) represents system of $n$ linear equations or the stationary Kirchhoff equation (\ref{ptp}), of  $n$ point vortices with identical circulation strength of Hydrodynamics \cite{aref},  with background flow $W(\bar{z_j})=\frac{1}{4l_B^2}\bar{z_j}$. These system of linear equations are same as the Berry connection defined in equation (\ref{b1}). We also show that the solution to these system of $n$ linear equations is the 1D-complex harmonic oscillator through supersymmetric quantum mechanics techniques. The Berry connection is associated with quantum momentum function of quantum Hamilton Jacobi formalism which allows to formulate the problem of quantisation in terms of topology, as the quantum numbers are topological invariant which arise due to singularities in the quantum momentum function. Quantisation arises as it continuously connects the topologically inequivalent Hamiltonians in the Hilbert space. 

We start with the equation of motion for  $n$ point vortices with identical circulation strength of Hydrodynamics \cite{aref}, which are known by the  Kirchhoff equation
\begin{equation}
\dot{z}_\alpha=\sum_{1\leq \alpha\leq N,\alpha\neq \beta}\frac{i\Gamma}{z_\alpha-z_\beta}-i\Omega\bar{z}_\alpha. \label{ptp}
\end{equation}
 $z_\alpha=x_\alpha+iy_\alpha$ are the position of $n$ vortices  and $\Gamma$ is the circulation or strength.   
It is clear, that the equation (\ref{ptp}) is identical to equation (\ref{b1}), with vortex strength $\nu$ and background flow $W(\bar{z}_\beta)=\frac{\bar{z}_j}{4l_B^2}$ with 
$\Omega=\frac{\bar{z}_j}{4l_B^2}$. The connection between the Laughlin wave function and the $n$ point vortices of Hydrodynamics is well studied problem in the literature \cite{pb1,pb2,pb3,aga}. In particular, the reference \cite{pb1} the Laughlin equation is obtained through the phenomenological approach by  quantization of Kirchhoff equation for $n$ point vortices, given in equation (\ref{ptp})
and the quantization condition is given by 
\begin{equation}
[z_\alpha,\bar{z}_\beta]=2l^2\delta_{\alpha\beta}.\label{com2}
\end{equation}

The paper is presented as following, in the section II the derivation of obtaining the Kirchhoff equation from the Schr\"odinger  equation is presented, in section III we give the self adjoint model Hamiltonian for fractonal quantum hall effect is given followed by discussion in section IV and finally conclude the paper in section IV.

\section{Schr\"odinger  equation and Kirchhoff equation}
In this section we show that the Schr\"odinger  equation is transformed to the system on $n$ linear equation, known as the Kirchhoff equation which describes the evolution of $n$ point vortices in Hydrodynamics. Consider the time dependent Schr\"odinger  equation with potential $V(x)=0$ gives
\begin{equation}
i {\frac {\partial }{\partial t}}\psi (x ,t)=\Gamma\frac{\partial^2}{\partial x^2}\psi (x ,t),\label{ts}
\end{equation}
where $\Gamma={\frac {-\hbar }{2m }}$.
By introducing a polynomial
\begin{eqnarray}
\psi(x,t)&=&(x-x_1(t))(x-x_2(t))\cdots (x-x_n(t))\nonumber\\&=&\prod_{k=1}^n(x-x_k(t)),\label{poly}
\end{eqnarray}
for $n=2$ substituting equation (\ref{poly}) in time dependent Schr\"odinger  equation (\ref{ts}) gives
\begin{equation}
-i\dot{x}_1(x-x_1(t))-i\dot{x}_2(x-x_2(t))=2\Gamma
\end{equation}
The equations considered at $x=x_1$ and $x=x_2$
\begin{equation}
\dot{x}_1=\frac{2\Gamma i}{(x_1-x_2)},\;\;\;\;
\dot{x}_2=\frac{2\Gamma i}{(x_2-x_1)}.
\end{equation}
By following a similarly for $n=3$, the equations at points $x=x_1$, $x=x_2$ and $x=x_3$ are given by
\begin{eqnarray}
\dot{x}_1&=&2\Gamma i\left[\frac{1}{(x_1-x_2)}+\frac{1}{(x_1-x_3)}\right]\nonumber\\
\dot{x}_2&=&2\Gamma i\left[\frac{1}{(x_2-x_1)}+\frac{1}{(x_2-x_3)}\right]\nonumber\\
\dot{x}_3&=&2\Gamma i\left[\frac{1}{(x_3-x_1)}+\frac{1}{(x_3-x_2)}\right]\nonumber
\end{eqnarray}
The same procedure for $n$ zeros gives
\begin{eqnarray}
\dot{x}_i=2\Gamma i\sum_{i\neq j}^n\frac{1}{(x_i-x_j)}.\label{kir1}
\end{eqnarray}
The equation (\ref{kir1}) are known as  the Kirchhoff equation which describe the evolution of $n$ point vortices in Hydrodynamics in the year (1887).
 Therefore, Schr\"odinger equation can be written in the form of system of $n$ linear equations (\ref{kir1}). The Kirchhoff equation (\ref{kir1}) with the background flow
  $\mathcal{W}(x)$ are given by
\begin{eqnarray}
\dot{x}_i=2\Gamma i\sum_{i\neq j}^n\frac{1}{(x_i-x_j)} +i\mathcal{W}(x).\label{kir}
\end{eqnarray}
 The solution to the stationary Kirchhoff equation (\ref{kir}), that is $\dot{x}_i=0$, is by found by Stieltjes electrostatic model \cite{st,st1}. In Stieltjes electrostatic model, there are $n$ unit moving charges between two fixed charges $p$ and $q$ at $-1$ and $1$ respectively on a real line and it is shown by Stieltjes that system attains equilibrium at the zeros of the Jacobi polynomials. It has been shown by the author, Stieltjes electrostatic model is analogous to the the  quantum momentum function of  quantum Hamilton Jacobi \cite{kvs} where moving unit charges are replaced by moving poles are like imaginary charges with $i\hbar$ placed between to fixed poles are like fixed charges.  In the process the background flow $\mathcal{W}(x)$ is identified with the superpotential.

In supersymmetry, the superpotential $\mathcal{W}(x)$ is defined in terms of the intertwining operators $ \hat{A}$ and $\hat{A}^{\dagger}$  as
\begin{equation}
  \hat{A} = \frac{d}{dx} + \mathcal{W}(x), \qquad \hat{A}^{\dagger} = - \frac{d}{dx} + \mathcal{W}(x).
\label{eq:A}
\end{equation}
This allows one to define a pair of factorized Hamiltonians $H^{\pm}$ as
\begin{eqnarray}
   H^{+} &=& 	\hat{A}^{\dagger} \hat{A} 	= - \frac{d^2}{dx^2} + V^{+}(x) - E, \label{vp}\\
  H^{-} &=& 	\hat{A}  {\hat A}^{\dagger} 	= - \frac{d^2}{dx^2} + V^{-}(x) - E, \label{vm}
\end{eqnarray}
where $E$ is the factorization energy. The partner potentials $V^{\pm}(x)$ are related to $\mathcal{W}(x)$ by 
\begin{equation}
  V^{\pm}(x) = \mathcal{W}^2(x) \mp \mathcal{W}'(x) + E, \label{gh}
\end{equation}
where  prime denotes differentiation with respect to $x$.

 As an illustration, we solve the Harmonic oscillator  in natural units using the Kirchhoff equation 
 \begin{equation}
 \sum_{1\leq j\leq n,j\neq k}^n\frac{1}{x_k-x_j}- x_j=0,\label{ufip}
\end{equation}
where $\mathcal{W}(x)=x_j$  is the superpotential of  Harmonic oscillator 
by introducing a polynomial
\begin{eqnarray}
f(x)=(x-x_1)(x-x_2)\cdots (x-x_n),\label{poly1}
\end{eqnarray}
and taking the  limit $x\rightarrow x_j$ and using l'Hospital rule one gets
\begin{eqnarray}
\sum_{1\leq j\leq n,j\neq k} \frac{1}{x_j-x_k}&=&\lim_{x\rightarrow x_j}
\left[\frac{f'(x)}{f(x)}-\frac{1}{x-x_j}\right]\nonumber\\&=&\lim_{x\rightarrow x_j}\frac{(x-x_j)f'(x)-f(x)}{(x-x_j)f(x)}
\nonumber \\&=& \frac{f''(x_j)}{2f'(x_j)}.\label{id}
\end{eqnarray}
By substituting the equation (\ref{id}) in equation (\ref{ufip}), we obtain 
\begin{equation}
f''(x_j)+2 x_jf'(x_j)=0.\label{poky}
\end{equation}
Hence the equation (\ref{poky}) is a polynomial, 
\begin{equation}
f''(x)+ 2xf'(x)=0.\label{pokyk}
\end{equation}
of order $n$ and is proportional to f((x) which gives Hermite differential equation
\begin{equation}
f''(x)+ 2xf'(x)+nf(x)=0.\label{pokyj}
\end{equation}
The readers should note that when we solve the problem for general potential say $Q(x_j)$ through Stieltjes electrostatic model one ends up with the following polynomial solutions
\begin{equation}
f''(x_j)+Q(x_j)f'(x_j)=0.\label{pokyl}
\end{equation}
the differential equation will have classical orthogonal polynomial solution only  when $Q(x_j)=\mathcal{W}(x_j)$ is superpotential.  As an illustration we the super-potential for the Coulomb potential is 
\begin{equation}
\mathcal{W}_{coul}(r_j)=\frac{1}{2}-\frac{(l+1)}{r_j}.
\end{equation}
Then by following  Stieltjes electrostatic model one gets
\begin{equation}
\frac{f''(r_j)}{2f'(r_j)}-(\frac{1}{2}-\frac{(l+1)}{r_j})=0,
\end{equation}
the polynomial
\begin{equation}
rf''(r)+(2(l+1)-r)f'(r)=0,
\end{equation}
 is proportional to $f(r)$  gives the Laguerre  differential equation. Similarly for Jacobi one gets
\begin{equation}
-\frac{f''(x_k)}{2f'(x_k)}-\frac{p}{x_k-1}-\frac{q}{x_k+1}=0.\label{pok}
\end{equation}
where the superpotential is given by
\begin{equation}
\mathcal{W}_{coul}(x_k)=\frac{p}{x_k-1}+\frac{q}{x_k+1}.
\end{equation}
the polynomial
\begin{eqnarray}
(1-x^2)f''(x)+2[q - p - (p + q)x]f'(x)=0.\label{poi}
\end{eqnarray} 
is proportional to $f(x)$  gives the Jacobi differential equation.
Therefore, one can solve all the bound state problems by Stieltjes electrostatic model
with the help of supersymmetric quantum mechanics.
\section{Laughlin Wave Function}

Consider the logarithm of Laughlin wave function (\ref{lf}) then differentiating  with respect to $z$ and equating to zero gives Kirchhoff equation
\begin{eqnarray}
i\frac{d}{dz_j}ln \psi(z_j)=\sum_{1\leq i\leq N,i\neq j}\frac{in}{z_i-z_j}-i\frac{1}{4l_B^2}\bar{z}_j =0.\label{kri}
\end{eqnarray}
The equation (\ref{kri}) is  Berry's Connection (\ref{b1}) for the positions of the quasi holes and the $\nu$ are the filling numbers defined in equation (\ref{lf}) and  similar by taking the transposition of  equation (\ref{kri}) we obtain equation for $A(\bar{z_j})$
\begin{eqnarray}
\sum_{1\leq i\leq N,i\neq j}\frac{in}{\bar{z}_i-\bar{z}_j}-i\frac{1}{4l_B^2}z_j =0.\label{kri1}
\end{eqnarray}
which describes the adiabatic transport of quasihole  at $z_i$ whan all other quasihole positions are fixed.
Therefore, from equation (\ref{kri})  it is clear that the Laughlin wave function (\ref{lf}) represents a Hamiltonian and the Kirchhoff equation are obtained by taking the minimum  of logarithm of the Laughlin wave function  (\ref{lf}), which represents  time independent Schr\"odinger  equation  with background flow $W(\bar{z_j})=\frac{1}{4l_B^2}\bar{z_j}$. It should be noted the stationary Kirchhoff equation (\ref{kri}) two dimensional and the Kirchhoff  equation (\ref{kir}) are one dimensional.

The equations (\ref{kri}) and  (\ref{kri1}) are Schr\"odinger  equations and  it immediately follows that the Hamiltonian is not hermitian as the superpotential is complex in nature. In other words, the Kirchhoff equations  are function of holomorphic coordinates $z=x+iy$ and the corresponding momenta 
 $\partial_{z_i} = \frac{1}{2}(\partial_{x_i}-i\partial_{y_i})$ and antiholomorphic  coordinates $\bar{z}=x-iy$ and the corresponding momenta 
 $\partial_{\bar{z}_i} = \frac{1}{2}(\partial_{x_i}+i\partial_{y_i})$. 
If one notices the Laughlin wave function (\ref{lf}) is a product of   the van der Monde’s determinant $\prod_{N \geq \beta > \alpha \geq 1}\left( z_\alpha-z_\beta \right)$ times the Gaussian  weight function
$\exp\left( -\sum_{\beta=1}^n  \frac{1}{4l_B^2} \vert z\vert_\beta ^2 \right)$.  The van der Monde’s determinant is a function of holomorphic coordinates and the the Gaussian  weight function is a function of holomorphic  and antiholomorphic coordinates. In literature these kind of function are studied in Segal-Bargmann space.  

To obtain the solution for the Kirchhoff equations (\ref{kri}) 
one should construct a self adjoint operator. Therefore, to construct a self adjoint operator we use the fact that lowest Landau levels are described in Bargmann space \cite{aga}. Hence, the solution to the Kirchhoff equations (\ref{kri}) 
as a self adjoint operator should be represented in terms of Gaussian  weight function
$\exp\left( -\sum_{\beta=1}^n  \frac{1}{4l_B^2} \vert z\vert_\beta ^2 \right)$.

By comparing the equations (\ref{kri}) and (\ref{ufip}) one can see that both of them represent the Harmonic oscillator, since the superpotential in equations (\ref{kri}) is complex in nature it should represent the 1D complex Harmonic oscillator. By following the procedure of supersymmeric quantum mechanics and  noting that the form of  superpotential corresponds to the position of the fixed poles or fixed charge for the given potential from Stieltjes electrostatic model. Therefore, for oscillator the fixed poles are at $\pm\infty$, hence, the superpotential is given by $\Omega\bar{z}$. Then the intertwining operators $\hat{A}$ and 
 $\hat{A}^{\dagger}$ are defined in equation (\ref{eq:A}) will read as
 \begin{equation}
 \hat{A}^{\dagger}=(-\partial_z+\Omega\bar{z}),\;\;\;\;\;\;\;\;\; \hat{A}=(\partial_{\bar{z}}+\Omega z),
\end{equation}  
then the Hamiltonian reads as
  \begin{equation}
H\psi(\vert z\vert)=\hat{A}^{\dagger}\hat{A} \psi(\vert z\vert)=(-\partial_z+\Omega\bar{z}) (\partial_{\bar{z}}
+\Omega z)\psi(\vert z\vert),\label{ha1}
  \end{equation}
 which gives 
  \begin{eqnarray}
  (\partial^2_{\vert z\vert}
 +\Omega\bar{z}\partial_{\bar{z}} -\Omega z\partial_z
  +\Omega^2\vert z\vert^2)\psi(\vert z\vert)=0.\label{dlu}
  \end{eqnarray}
The equation (\ref{dlu}) is the complex Hermite polynomial differential equation \cite{chp} and their solutions are given by 
 \begin{equation}
\psi (\vert z\vert)={\frac {1}{\sqrt {2^{n}\,n!}}}\cdot \left(\Omega\right)^{1/2}\cdot e^{-\Omega^2\vert z\vert^2}\cdot H_{n}\left(\Omega\vert z\vert\right), n=0,1,2,\ldots\label{ch}
\end{equation}
The operator $H$, given in equation (\ref{ha1}), is also known as Landau Hamiltonian  which describes  a charged nonrelativistic particle moving under the action of an external uniform magnetic field  applied perpendicularly  \cite{chp,chp1,chp2,chp3,chp4}
 \begin{eqnarray}
H\psi(\vert z\vert)=4  (\partial^2_{ z\bar{z}}
 +B\bar{z}\partial_{\bar{z}} -B z\partial_z
  +B^2\vert z\vert^2)\psi(\vert z\vert).\label{dlui}
  \end{eqnarray}
where $B$ is magnetic field gives the Landau levels $4B(2l+1)$ \cite{chp4}. 
The magnetic Schr\"odinger operator
$H$, associated to the vector potential $ A=(2By ,−2Bx,0)$, given in $x$, $y$ coordinates
\begin{eqnarray}
H =  -((\partial_{ x}+iBy)^2 + (\partial_{ y}-iBx)^2)
 \label{dluc}
  \end{eqnarray}
  This Hamiltonian is identical to the model Hamiltonian proposed by Handane \cite{hal} and also well studied im mathematical Literature \cite{chp5}. The solution to the self adjoint operator (\ref{dlui}) are given by complex Hermite polynomials (\ref{ch}).

\section{Discussion}
It is well known by applying the Cole-Hope transformation $\psi=e^{iS}$, where $\psi$ is wave function and $S$ is characteristic function, on the Schr\"odinger  equation  and then by taking in the limit $\hbar\rightarrow 0$, reduces to classical Hamilton Jacobi equation. In the classical Hamilton Jacobi, equation of motion is governed by the Hamiltonian and the problem is solved by continuously transforming the Hamiltonian from the initial state to the final state through canonical transformations. In quantum mechanics, the equation of motion is governed by the Schr\"odinger equation which is described by a Hamiltonian. It is not possible to continuously transform the Schr\"odinger equation from the initial state to the final state through a canonical transformation as it has singularities  and it is quite evident when the Schr\"odinger equation is written in terms of  Kirchhoff equations (\ref{kir}).  The analogy $n$ point vortices of Hydrodynamics with the fractional quantum Hall effect has allowed us to identify  the Schr\"odinger equation  as Berry connection. Thus, by making the Berry connection exact gives rise to quantisation. Hence, the quantum numbers are topological invariants arising due to singularities in the Schr\"odinger equation . Therefore, we conclude that  quantisation  arises as it continuously connects the topologically inequivalent Hamiltonians in the Hilbert space. 
\section{Conclution}
In conclusion, we shown that the Laughlin wave function represents a Hamiltonian and  the  Berry connection associated with Laughlin wave function is the Schr\"odinger  equation. We have also proposed the solution for the Laughlin wave function are 1D complex Hermite polynomials. The analogy $n$ point vortices of Hydrodynamics with the fractional quantum Hall effect has allowed us to identify  the Schr\"odinger equation  as Berry connection. This allowed us to formulate the problem of quantisation in terms of topology, as the quantum numbers are topological invariant which arise due to singularities in the  Kirchhoff equation. Therefore, Quantisation arises as it continuously connects the topologically inequivalent Hamiltonians in the Hilbert space.


\begin{thebibliography}{99}
\bibitem{pd}  Konig, Markus et.al, Science. 318 (5851): 766–770. (2007)
\bibitem{qh}K. V. Klitzing; G. Dorda; M. Pepper .  Physical Review Letters. 45 : 494–497,(1980).
\bibitem{fqh}Tsui. D. C; Stormer H. L;  Gossard A. C., Physical Review Letters. 48 : 1559,(1982). 
\bibitem{rbl} R. B. Laughlin, Phys. Rev. Lett. 50, 1395
\bibitem{dt} David Tong, 
Lecture notes on quantum Hall effect, http://www.damtp.cam.ac.uk/user/tong/qhe.
\bibitem{kvs} K. V. S. Shiv Chaitanya,  Pramana journal of physics, Vol. 83, No. 1, 139, 2014.
\bibitem{aref} Hassan Aref, Ann. Rev. Fluid Mech, 15:345-89, (1983).
\bibitem{st} T.J. Stieltjes, , Comptes Rendus de 
l'Academie des
Sciences, Paris, 100 (1885), 439-440; Oeuvres Completes, Vol. 1, 440-441.;

\bibitem{st1}T.J. Stieltjes, Comptes Rendus de l'Academie des
Sciences, Paris, 100 (1885), 620-622; Oeuvres Completes, Vol. 1, 442-444.
\bibitem{met} Mehta, M.L. (2004). Random Matrices, 3rd Edition, Pure and Applied Mathematics
(Amsterdam), 142, Amsterdam, Netherlands: Elsevier/Academic Press.
\bibitem{pb1} P. Wiegmann, Journal of Experimental and Theoretical Physics, Vol 117, issue 3, 538, (2013).
\bibitem{pb2} P. B. Wiegmann,  Phys. Rev. B 88, 241305(R)(2013)
\bibitem{aga} Alexander G Abanov, J. Phys. A: Math. Theor. 46 292001, (2013).
\bibitem{pb3} Paul Wiegmann and Alexander G. Abanov
Phys. Rev. Lett. 113, 034501, (2014)
\bibitem{hal} Haldane, Phys. Rev. Lett. 51, 605 (1983).
\bibitem{chp} Allal Ghanmi,J. Math. Anal. Appl. 340, 1395, (2008).
\bibitem{chp1}A. Intissar, A. Intissar,  J. Math. Anal. Appl. 313 (2) (2006) 400–418.
\bibitem{chp2} I. Shigekawa, J. Funct. Anal. 75 (1)
 92, (1987).
\bibitem{chp3} S. Thangavelu, Lectures on Hermite and Laguerre Expansions, Math. Notes, vol. 42, Princeton Univ. Press, Princeton, 1993.
\bibitem{chp4} Allal Ghanmia and Ahmed Intissar, JMP, 46, 032107 (2005)
\bibitem{chp5} J. Avron, I. Herbst, and B. Simon,  Duke Mathematical Journal, vol. 45, no. 4, pp. 847–883, (1978).
 \end{thebibliography}
  \end{document}